%% file: main.tex
\documentclass[11pt]{article}
\usepackage{amsfonts}
\usepackage[T1]{fontenc}
\usepackage{amssymb,amsmath}
\usepackage{mathrsfs}
\usepackage{amsthm}
\usepackage{fancyhdr}
\usepackage{times}
\usepackage{color}
\usepackage[usenames,dvipsnames,svgnames,table]{xcolor}
\usepackage{textcomp}
\usepackage{hyperref}
\usepackage{graphicx}
\usepackage{caption}
\usepackage{subcaption}
\usepackage{float}
\usepackage{enumitem}
\usepackage{graphicx}
\usepackage{fullpage}
\usepackage{subcaption}

\usepackage{authblk}
\usepackage{blindtext}

\begin{document}

\title{Clustering high dimensional meteorological scenarios: results and performance index}
\author[1]{Yamila Barrera}
\author[2]{Leonardo Boechi}
\author[2]{Matthieu Jonckheere}
\author[3]{Vincent Lefieux}
\author[4]{Dominique Picard}
\author[1,5]{Ezequiel Smucler}
\author[1]{Agust\'in Somacal}
\author[1]{Alfredo Umfurer}
\affil[1]{Aristas S.R.L., Dorrego 1940, Torre A, 2do Piso, dpto. N (1425), CABA, Argentina}
\affil[2]{Instituto de Calculo-CONICET, Intendente Guiraldes 2160, Ciudad Universitaria, Pabell\'on II, 2do. piso, (C1428EGA), Buenos Aires, Argentina}
\affil[3]{R\'eseau de Transport d' Electricit\'e(RTE), Paris, France. }
\affil[4]{Universit\'e de Paris, LPSM, UFR Mathematiques Batiment Sophie Germain, 
75013 Paris, France.}
\affil[5]{Universidad Torcuato Di Tella,  Av. Figueroa Alcorta 7350 (C1428BCW) S\'aenz Valiente 1010 (C1428BIJ) CABA, Argentina.}

\affil[ ]{\textit {\{y.barrera, e.smucler, a.somacal, a.umfurer\}@aristas.com.ar}}
\affil[ ]{\textit {lboechi@ic.fcen.uba.ar, m.jonckhe@dm.uba.ar, vincent.lefieux@rte-france.com, picard@math.univ-paris-diderot.fr}}


\date{}
\maketitle


\begin{abstract}
The Reseau de Transport d'Electricit\'e (RTE) is the French main electricity network operational manager and dedicates large number of resources and efforts towards understanding climate time series data. 
We discuss here the problem and the methodology of grouping and selecting representatives of possible climate scenarios among a large number of climate simulations provided by RTE.
The data used is composed of temperature times series for 200 different possible scenarios on a grid of geographical locations in France. These should be clustered in order to detect common patterns regarding temperatures curves and help to choose representative scenarios for network simulations, which in turn can be used for energy optimisation. 
We first show that the choice of the distance used for the clustering has a
strong impact on the meaning of the results: depending
on the type of distance used, either spatial or temporal patterns prevail.
Then we discuss the difficulty of fine-tuning the distance choice (combined with a dimension reduction procedure) and we propose a methodology based on a carefully designed index.
\end{abstract}


\section{Introduction}
\label{sec:intro}
\input{secs/intro.tex}

\section{Meteorological data}
\label{sec:data}
\input{secs/data.tex}

\section{Methodology}

\label{sec:clustering}
\input{secs/clustering_techniques.tex}

\subsection{Evaluation index}
\label{sec:index2}
\input{secs/index2.tex}


\section{Results}
\subsection{Significance of classical distances}
\label{sec:dist}
\input{secs/dist.tex}

\input{secs/experiments.tex}

\section{Conclusion}
\label{sec:conclusion}
\input{secs/conclusion.tex}


\bibliography{main.bib}   
\bibliographystyle{apalike}       


\end{document}

%% file: secs/intro.tex
Temperature fluctuations have a strong influence on the electric consumption. 
As a consequence, identifying and finding groups of possible climate scenarios is useful for the analysis of mitigation and adaptation policies of the electric supply system \cite{climatechangechina}. 
Given a set of time series representing temperature, a climate segmentation can potentially lead to a simplification of the posterior energy consumption analysis. 

The work done by \cite{climateRegions} focused on finding homogeneous climate regions in the French territory, that is, finding groups of geographical points with similar temperature time series. 
Without explicitly including in the model the spatial relationships between the time series, they were able to recover the French territory regions.
Other authors, like  \cite{unitedstates} for the USA climate types and \cite{toulouse} for Toulouse have used different clustering techniques for the same purpose.
A methodology for classifying the land surface into climate types using several climate variables, not only temperature series, was proposed by \cite{global_climate_clustering}.


Clustering is an unsupervised learning technique aiming at finding patterns (usually groups) in unlabeled data set.
In the present case, clustering algorithms will be used to find homogeneous (in a sense to be defined carefully) groups of times series.
As it is well known, the dimensionality of the data can be a formidable obstacle to the clustering techniques, due in great part to the phenomenon of concentration of measures in high dimension \cite{high}: as the dimension grows, the usual distance (say L2) between two uniformly chosen data points concentrates around a constant value, making the identification of groups much more difficult.
As a consequence, the main challenge when using clustering algorithms is to jointly choose a distance measure and a dimension reduction technique to break the concentration of distances. In the case of time series, the dimension (being the number of time points) is naturally high and
these choices become particularly crucial.

On the other hand, given the spatio-temporal nature of our data set,
the choice of the distance has strong implications in terms
of "feature selections", i.e. on the specific patterns that will influence the clustering results. As it turns out,  a detailed study of the choice of the distance effects on the data set needs to be undertaken. Otherwise, the final clustering results could be of little relevance for the practitioner.

\subsection*{Related work}
Several recent review papers give a good account of the important body of literature on time series clustering in the last two decades (see for instance \cite{review1}).
On the one hand, there has been interesting work on finding new distances
tailored to compare time-series (e.g. \cite{kshape, DTW}), which allow
to adapt clustering techniques to the particularities of time-ordered signals.
Such distances have been showed to be relevant for many use cases where interesting patterns are present at an intermediate time scale, see e.g. \cite{kshapeE}.
However, there is generally no clear consensus regarding the best performing one, as it seems to depend on the domain of application.

On the other hand, many recent papers deal with the notion of deep clustering (see \cite{DC2,DC3} and the survey \cite{DCsurvey}) where the idea is to mimic a supervised methodology
by choosing jointly a transformation of the data and a clustering procedure to minimize a loss function. 
This line of research has been shown to have very high performance e.g. in some imaging, computer vision and spatio-temporal applications \cite{DC1}. 
However, as with many deep clustering techniques, 
a central issue might be the interpretability of the results and some difficult fine-tunning might be necessary to find interesting patterns \cite{DCsurvey}. These issues become critical in our setting where the basic ingredients (distance and data transformation allowed) might distort completely the results. Though a promising research direction, we believe that many elements (in particular the  impact of the distance and the dimension reduction) have to been studied first to obtain efficient and meaningful results.
As a consequence, we decided to put our efforts in this first study on decoupling the data transformations and the clustering tasks, keeping a very simple clustering method and using an interpretable index to evaluate jointly both objectives to gain understanding on the effect of varying distances and transformations.

\subsection*{Data and objectives}
The data that we are considering in this work are time series of hourly measured temperatures  over a grid of geographical points in France and neighboring areas. These series have data for 200 years, each of which can be considered a different climate scenario. 
The R\'eseau de Transport d'Electricit\'e (RTE) is the electricity transmission system operator of France. RTE, as other energy players have for many years expressed their needs for long series of data representative of the climate, scanning a maximum of possible hazards.
Constant climate scenarios should be interpreted as sets of possible achievements of 200 years under the same climate.
These are neither re-analyzes of past situations nor forecasts. Long simulated climate data series provide a vast sample of meteorological situations.
In other words, for each geographical location, 200 temperature time series representing 200 temperature scenarios are available. 
On the one hand it is difficult to pretend that all these scenarios appear with equal probability. 
On the other hand, there is no real tool to have a precise idea on the probability of each scenario. 
Hence a more realistic technological advance towards this evaluation is to find groups of scenarios.

\subsection*{Contribution}

Our task consists both in finding clusters of interest for the experts, 
 but also in explaining in which sense they are appropriate.
Observe that we actually deal with an intrinsically multi-objective problem where 
the efficiency of the clustering algorithm (in terms of, e.g., minimising the intra-cluster distances and maximizing the inter-cluster distances)
has to be considered jointly with a penalization on the
dimension reduction. Indeed, as it is intuitive, a drastic dimension might strongly deteriorate the meaningfulness of the final results. We will further show that even 
a mild dimension reduction might have far-reaching consequences.


Our main contribution is then two-fold:
\begin{itemize}

\item
On the one hand, we give 
an interpretation {\bf for our data set} of the several types of distances / transformations used in the time series literature.
We show in particular that there is a quite counter-intuitive phase transition when the amount of dimension reduction grows, in terms of spatial vs temporal effects on the distance between time-series.
In other words, when drastically reducing the dimension, only the temporal effects
(different scenarios associated to a given place) are dominating. When keeping the 
same amount of information, the spatial differences are dominating.
Hence, depending on the subsequent purpose of the clustering results (e.g., defining climate homogeneous regions or defining representative scenarios), specific distances should be carefully picked and employed.

We believe that
such conclusions are, to a certain extent, generalizable or at least meaningful to other data sets of time series.

\item
On the other hand, we actually perform clustering of climate scenarios using several clustering pipelines including choices of distances, transformation of the data and clustering algorithms. 
 Apart from providing a partition into groups, we also provide in each group a ``typical scenario''. This could be a very useful tool for expert interpretation of each specific group.
 
The fine-tuning of the choice of the distance/dimension reduction is 
done using a tailored-made index based on embedding ideas, and allowing to reach an appropriate trade-off between clustering efficiency (where dimension reduction plays a positive role) and data distorsion (where dimension reduction can play a negative role). This allows to partially validate our clustering results,
in parallel of the validation provided by RTE experts.
When the focus is put on clustering one-year-scenarios, we show that the best results are reached with the lagged correlation distance and to a minor extent with DTW.

\end{itemize}

The paper is organized as follows.
Section \ref{sec:data} gives a description of the data provided by Meteo-France through RTE.
Section \ref{sec:clustering} gives a review of the clustering techniques used in the paper, including data transformation, distance measures and clustering algorithms.
 Section \ref{sec:dist} explains the qualitative properties of classical distances in our context and highlights a phase transitions in terms of spatial vs temporal explainability.
   Section \ref{sec:index2} defines our comparison index.
 Section \ref{sec:experiments} conducts the clustering experiments in the meteorological data and gives indication on the performance of our comparison index.
Section \ref{sec:conclusion} gives conclusions and future lines of work.

%% file: secs/data.tex
\label{meteorological.data}
\subsection{True data for forecasters and scenarios} 

Observation is the first step in a forecast. 90\% of the observation data used by Meteo-France's forecast models come from meteorological satellites. The remaining 10\% are provided by ground stations, radiosonde, airborne and commercial vessel sensors, or installed on anchored and drifting buoys. Meteo-France also receives other meteorological services from measurements collected around the globe.
All these observations are then processed to extract useful information  towards the forecasting model: one speaks of data assimilation. About 22 million observations data are used every day by the models at the end of the assimilation step. The data from the observations are combined with other information, such as very recent forecasts, to establish an initial state of the atmosphere that the model will be able to use.

From this initial state, the models simulate the evolution of the atmosphere, which they cut into a grid in three dimensions with meshes of different sizes. Simulation is based on the physical laws that govern atmospheric evolution: mainly the laws of the mechanics of fluids, supplemented by those governing changes in the state of water (condensation, evaporation, precipitation formation), turbulence, radiation or the many interactions with the Earth's surface and even space.

To describe the state of the atmosphere and to perform their calculations, the weather models cut the atmosphere into elementary boxes each containing a value of pressure, wind, temperature, humidity and others. On the horizontal, this decoupage is defined by the mesh distance of the grid of the model, and on the vertical, by the number of levels of the model. Mesh distance and number of levels vary according to the desired fineness, the computing power available, the vocation of the model (short-term forecast, climate simulation, seasonal forecast for instance).

For phenomena exceeding the size of their mesh, the models follow these physical laws in all rigor. On the other hand, smaller phenomena are not explicitly described in the model. They are taken into account by means of specific algorithms that simulate their average influence within the meshes of the model.

Note that the results of these simulations performed by the models are not  forecast but scenarios of evolution of the main meteorological parameters in all the points of the grid which represents the atmosphere.

 \subsection{Scenarios}
For points on a 76 x 51 grid, which includes all the French territory and surrounding areas, hourly measurements of temperature over a 200 year span are available, with each year representing a possible climate scenario.

The temperature measurements are available hourly over each geographical point and over a time span of 200 `years'. These daily observations are contiguous in that, for example, the last day of first year is followed by the first day of the second year. However, we will not use this contiguity information and each `year' will be interpreted as a different possible climate scenario. This is what we mean in the following, whenever we refer to a climate scenario.
Instead of studying the  time series on the whole year we also decided to restrict to winters, which are crucial for energy providers.

So, in the sequel, we are considering the observation of 200 time-series of hourly observed temperatures on 90 days of winter, all corresponding to the same geographical point. As a preprocessing, we
center each series  by the global mean (considering all geographical points, all scenarios).

%% file: secs/clustering_techniques.tex
For general clustering algorithms, four important choices enter into play :
\begin{enumerate}

\item The representation of the data in the feature space (which can be the original space) and the dimension reduction to transform the original data into features.

\item The distance used to measure dissimilarities between the representations of the data in the feature space and the distance used to measure dissimilarities between the original data.

\item The number of clusters.

\item The actual method to partition data points into clusters.
\end{enumerate}

In what follows, we discuss several alternatives considered for these choices.

\subsection{Representations and distances}

This section concerns the first two points listed above as key factors of a clustering algorithm, by exhibiting various techniques we considered to represent the data and to compute distances between the representations.

\subsubsection{Representations}
\label{sec:representations}

\subsubsection*{PCA}

Given a data set where each data point is in $\mathbb{R}^T$, the PCA algorithm produces a predetermined number of directions (the principal directions) on which to project the data points to best capture their variability. The resulting projections are the so called principal components.  The number of principal components is chosen to explain at least $95\%$ of the variability in the data.

\subsubsection*{Functional approximation featuring}
Another point of view consists in taking into account the fact that each data point is in fact a function of time observed at regular instances. 
We can represent $Y_t$, the temperature at a given point at time $t$ as a random variable of the form
$$Y_t=f(t\delta)+e_{t}, \quad t = 1,\ldots,T,$$ 
for some error random variable $e_{t}$ and some mesh $\delta$.
The function $f$ can then be estimated in a nonparametric fashion using a dictionary of basis functions, see for instance \cite{Tsybakov}.
We will use two main dictionaries, the Haar basis and the Fourier basis taking as assumptions that the function $f$ is sufficiently smooth (for the Haar basis) or has some prominent  frequencies such as day-night (in the case of the Fourier basis).
The number of terms in the basis expansion is chosen to explain at least $95\%$ of the variability in the data.
 
\subsubsection{Distances}

Any clustering algorithm requires a notion of distances between the objects being grouped.
This is actually the key ingredient of any clustering methodology.
In our case, two time series corresponding to different locations and/or scenarios will be deemed `similar' whenever their distance is `small'. Next, we list some of the notions of distance between time series that we considered useful.
Let $z=\left\lbrace z_{t}\right\rbrace_{t=1}^{T}$ and $w=\left\lbrace w_{t}\right\rbrace_{t=1}^{T}$ be any two time series.

\subsubsection*{Euclidean}

This is just the standard euclidean distance between two vectors, i.e.,  the distance is $\Vert z-w\Vert_{2}=\left(\sum_{t=1}^{T} (z_{t} - w_{t})^{2}\right)^{1/2}$. Note that this does not take into account the dynamic nature of the data at all, but has demonstrated to be competitive in terms of accuracy \cite{kshape}.
\\
Note that this euclidean distance is used in the data space or in the feature space when the data are represented by their projections (PCA, functional bases)  or their autoencoder representatives

\subsubsection*{Max Lagged Pearson Correlation (MLPC)}
The MLPC seeks for an optimal alignment between two signals, with the two series only being allowed to be aligned via shifts in the time axis.  The distance between two times series $z$ and $w$ is defined as:
$$SBD(z, w ) = 1-\max_{\vert k\vert\leq  k_{max}} {\text{corr}_{k}(z,w)} $$

where $\text{corr}_{k}$ is the Pearson correlation at $k$ lags and $k_{max}$ is a user-specified constant. 
The distance is always equal or greater than 0 and equal or less than 2. 

This comparison has good properties for signal presenting similarities except for a fixed translation in time.
This distance was inspired by the one proposed in \cite{kshape}, as part of the k-shape clustering algorithm.

\subsubsection*{Dynamic Time Warping}
 Introduced by \cite{DTW}, this distance essentially looks for the optimal, possibly non-linear, alignment between the two series. For details see \cite{DTW}. 
 \cite{CDTW} proposed a constrained version with similar performance results but less computational complexity.

Note that both DTW and MLPC, unlike the traditional Euclidean distance, take into account the dynamic nature of the data, the fact that we are dealing with time series, which by their very definition can have lagged relations. Unlike the Max Lagged Pearson-Correlation, DTW allows for non-linear alignments of the series and is computationally more demanding.

\begin{figure}[H]
\centering

\begin{subfigure}{.5\textwidth}
  \centering
  \includegraphics[width=\linewidth]{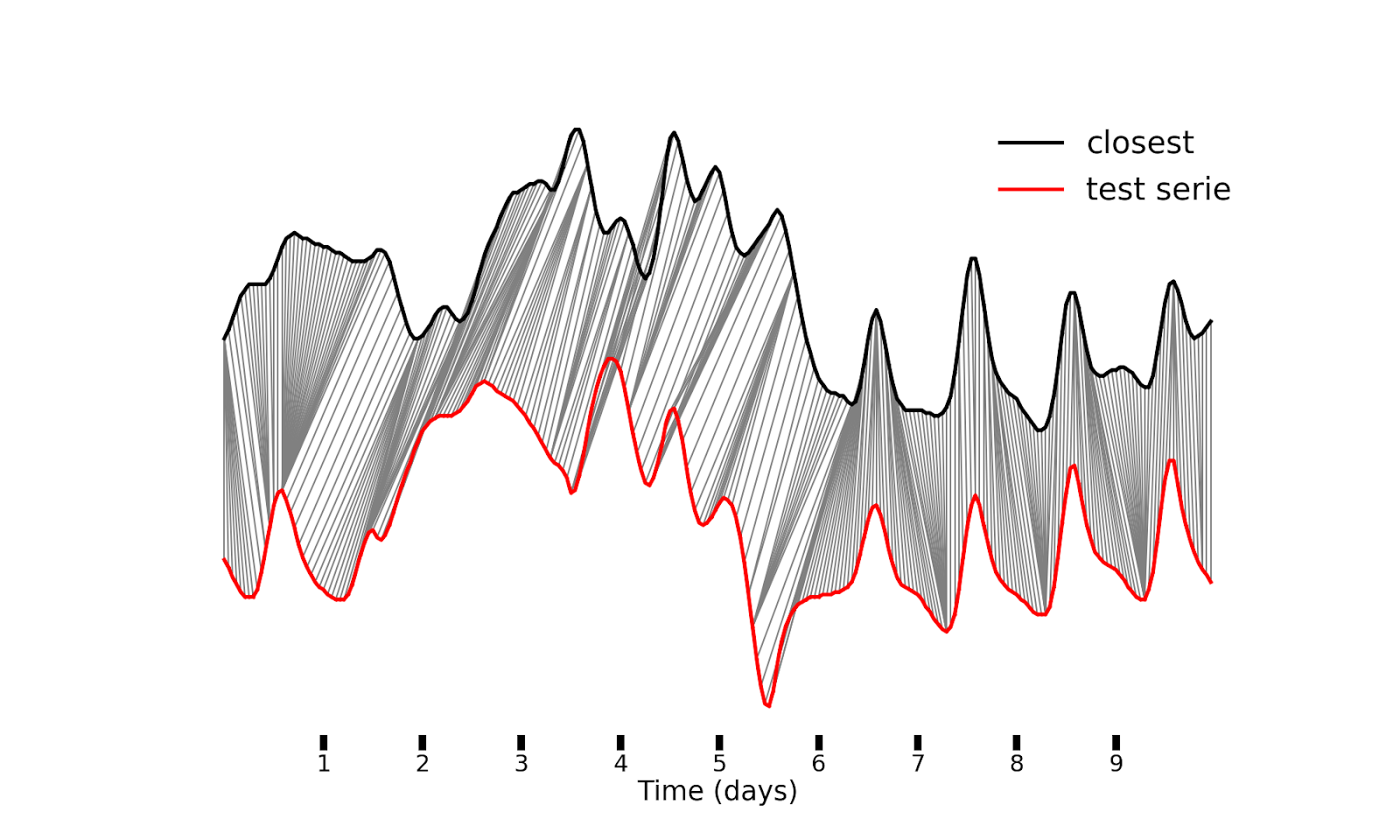}
\end{subfigure}%

\begin{subfigure}{.5\textwidth}
  \centering
  \includegraphics[width=\linewidth]{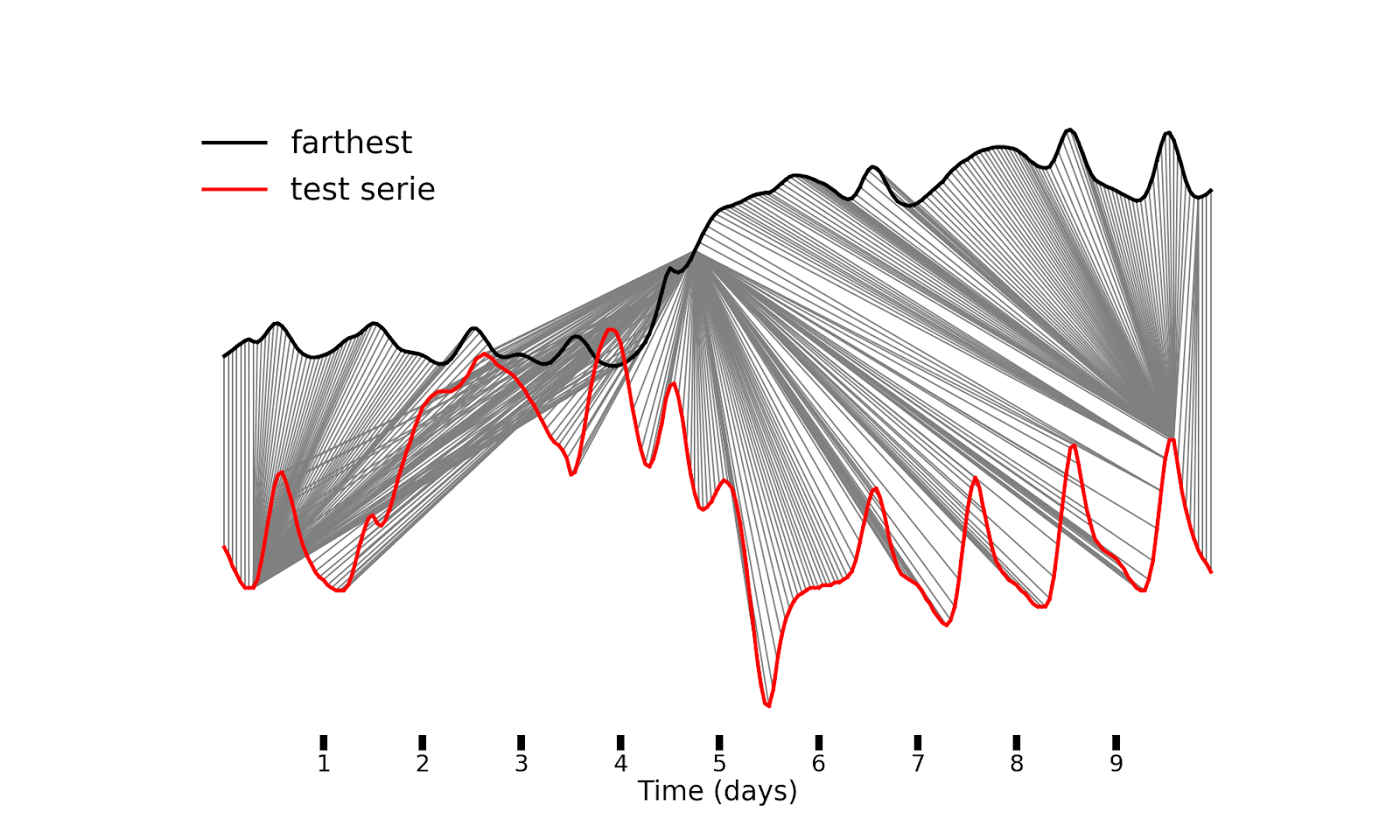}
\end{subfigure}

\caption{Left panel: test series in red and the series closest to it in black; grey lines show the optimal alignment. Right panel: test series in red and the series farthest from it in black; grey lines show the optimal alignment.}
\label{fig:DTW_examp}
\end{figure}

The plots in Figure \ref{fig:DTW_examp} illustrate how the DTW distance works. In the top panel we see a test series in red and the series closest to it in black; grey lines show the optimal alignment. In the bottom panel we see the test series in red and the series farthest from it in black; grey lines again show the optimal alignment. The series closest to the test series mainly differs in a small warping of the time axis. On the contrary, the series farthest from the test series has completely different dynamics. This is what the DTW distance seeks to capture. The following 3D interactive visualizations may help the rather gain further insights into DTW: \url{https://plot.ly/~aumfurer/2/closest}, \url{https://plot.ly/~aumfurer/4/farthest}.

We should mention that the added flexibility associated with DTW, namely, it allowing for non-linear alignments, comes with the price of a significant computational cost when compared to the Max Lagged Pearson-Correlation distance. As a matter of fact, the worst case complexity for computing DTW is $O(T^2)$, an order worse than the $O(k_{max} T)$ needed for the MLPC. Moreover, results obtained using the DTW distance can be much harder to interpret, due to that added flexibility.

\subsection{Clustering method: K-Medoids}
\label{subsec:kmedoids}
We now quickly present the simple clustering algorithms that is considered in our analysis. 
K-Medoids is a very commonly used clustering algorithm. 
It is an easily implementable and interpretable  procedure which  aims at producing a local minimum to the following problem: {\textit{ Given the number  $k$ of  clusters,}}  find a partition of the data points into $k$ disjoint clusters such that the sum of the distances of members of a cluster to a `cluster representative' is minimal. 
The optimisation problem is similar to K-means but unlike K-Means, cluster representatives in K-Medoids are  data points whereas in K-means they are barycenters. The algorithm uses random starts, so the result of one run of the algorithm is a local minimum depending on this random starts. As a consequence, the algorithm has to be run a significant number of times to get a global minimum. See \cite{clustering} for details.
In practice, it works as follows. The centroids are randomly initialized at some data points. Each point in the data-set is then assigned to the centroid closest to it. Then, the centroids in each cluster are updated to the cluster members that minimize the sum of the distances to all other members of that cluster. Finally, each point is re-assigned to the cluster with the centroid closest to it. These two last steps are repeated until convergence.

K-means and K-Medoids often give comparable results. K-means exactly optimizes the euclidean distance, K-Medoids has good properties of robustness with respect to outliers.

 We chose this clustering method among many others, because of various reasons : simplicity of implementation, control of the number of clusters (chosen in advance), statistical interpretation, stability (if the number of runs is important enough, and the number of clusters adequate) and the ease of  interpretability of the clusters (since they have one representative which is a real point and not a barycenter).

%% file: secs/index2.tex
As usual in unsupervised learning, in the absence of ground truth, there is no direct way to  detect results of poor quality i.e. far from the theoretical 'hypothetical' one (as opposed to estimating  a regression function for instance where the observed errors are giving hints).

In this section, we propose and explore a new evaluation methodology by defining an index measuring the trade-off between the faithfulness of the representation to the original data on the one hand and the clustering efficiency in the feature space on the other hand.
The difference with e.g. loss functions introduced in some literature on deep clustering \cite{DCsurvey} is the fact that this index can generalize to:
\begin{itemize}
\item any distances in the original and the features spaces,

\item any transformation from the original to the feature space,
(even when there is no well defined notion of best reconstruction).
\end{itemize}

Our index has two components and is trying to balance two opposed effects, standard in clustering : by reducing drastically the dimension we obtain a very stable clustering with  generally very good measures of efficiency, but quite poor interest and interpretability since the representation of the data in the feature space is weak.

\ 

{\textit{Faithfulness} Our index is inspired by the stochastic neighbors embeddings related techniques.
We start by defining a probability distribution $P$ using a Gaussian model for the embedding in the original space equipped with a basic distance (euclidean or k-shape or DTW) (see for instance \cite{tsne} for the exact  procedures).
Then, we define a probability distribution $Q$ on the feature space using a Gaussian distribution for the embedding since the dimension is expected to be still reasonably high except in some pathological cases.
The intuition here is that $P$ should represent a high-dimensional space
while $Q$ should represent the "projection" into a lower (but still reasonably high) dimensional space.

Following classical approaches, we then define the distance between our evaluation distribution $P$ and our feature distribution $Q$, as the symmetrized Kullback-Leibler:
$$ D_{JS} (P, Q) = D_{KL}(P||Q)/2 + D_{KL}(Q||P)/2.$$ 
Since this metric gives a value between 0 and infinity, we will use a logistic function to restrict to the interval $[0, 1)$.
Then our "Fidelity" measure $F(d)$ is defined as:
    $$
    F = {2 \over  1+ e^{ D_{JS}(P,Q)}}.
$$
The intuition here is that there should be some quite sharp transition (depending of the typical behavior of $D_{JS}$) between
faithful enough representations and unmeaningful ones. 

\

{\textit{Clustering efficiency and balance}}

\

On the other hand, we now define an index $W$ that we call
the within index 
 as an efficiency  measure for the clustering in the feature space. We use an adaptation of classical 
clustering indices (for instance the Carinski-Harabasz index), which is convenient to measure the efficiency of the $k$-medoid algorithm, as it corresponds to a normalized version of the functional optimized by the $k$-medoid algorithm and which is defined as follows.

For a given clustering and for any time series $z_i$, define $\mu(z_i)$ as the center of the cluster assign to $z_i$
by this clustering.

Then the within index is simply the normalized ratio of the mean squared distances of series to their centers with the mean square distance between series:
$$
W={ {1 \over N} \sum_{z_i} (d(z_i, \mu(z_i) )^2  \over   {1 \over N(N-1)} \sum_{z_i, z_j} (d(z_i,z_j))^2 } 
$$


\

Our final index combines in a multiplicative manner both measures:
$$I = F  (1-W).$$
This multiplicative structure (compared to more standard additive penalizations) is simply due to the exponential shape of our fidelity index.

%% file: secs/dist.tex
\begin{figure}[h]
    \centering
    \includegraphics[scale=0.4]{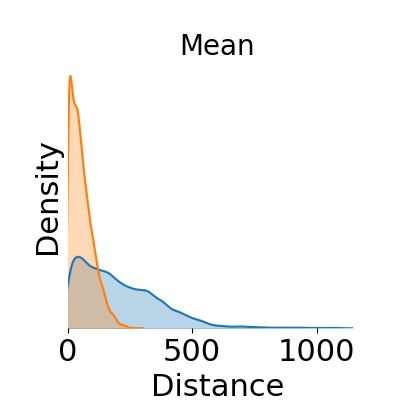}
    \includegraphics[scale=0.4]{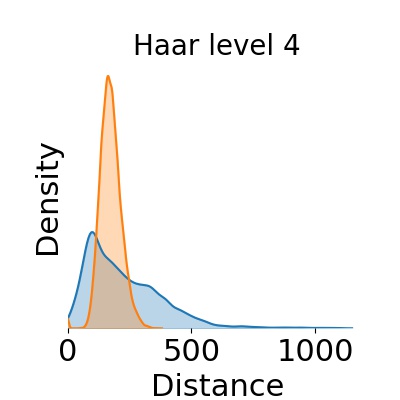}
    \includegraphics[scale=0.4]{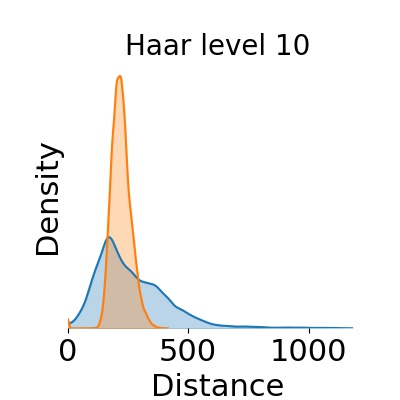}
    \includegraphics[scale=0.4]{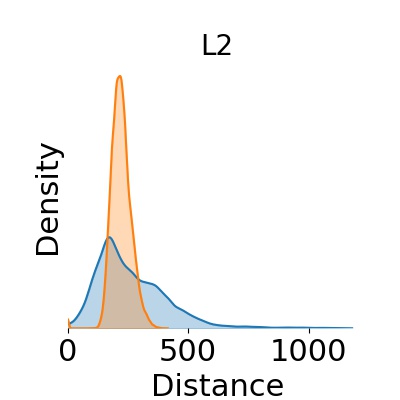}
    \includegraphics[scale=0.4]{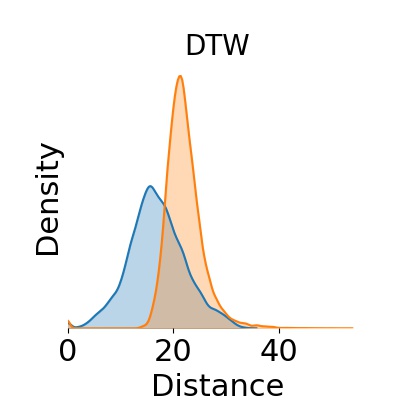}
    \includegraphics[scale=0.4]{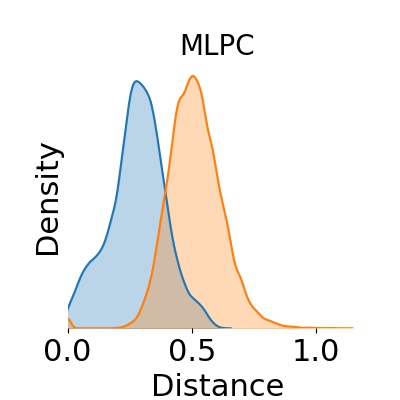}
    \caption{ Distance distributions when considering different years for a geographical point (orange) or different geographical points from same year (blue).}
    \label{fig1}
\end{figure}

To understand in depth the implication of the choice of the distance, we perform the following experiments.
We choose a given geographical point in France for a particular year as a reference. That is, our reference time series corresponds to the temperatures for a latitude and longitude point in France for a particular scenario. This time series is compared with : 
\begin{enumerate}[label=\Alph*]
    \item Times series corresponding to the same geographical point along other scenarios (there are 198 other scenarios different from the reference one).  
    \item Times series corresponding to other geographical points in France but the same scenario that the reference series (the geographical grid has more than 3500 points, but we compared it with 198 randomly chosen points so that the number of series in group B is the same than in group A).
\end{enumerate}
This analysis was repeated for different geographical points and scenarios as reference time series, and the results were similar. We hence show the results of just one reference temperature times series. 
We compute the distances in table \ref{table:distances} between the reference series and series from group A (orange in figure \ref{fig1}) and between the reference time series and series from group B (blue in figure \ref{fig1}). 

\begin{table}[h]
\centering
\begin{tabular}{|l|l|l|ll}
\cline{1-3}
\multicolumn{1}{|c|}{\textbf{Transformation}} & \multicolumn{1}{c|}{\textbf{Distance}} & \multicolumn{1}{c|}{\textbf{Name}} &  &  \\ \cline{1-3}
plain time series                                             & L2                                      & L2                                   &  &  \\ \cline{1-3}
Haar, energy 0.95                            & L2                                      & Haar95                              &  &  \\ \cline{1-3}
zscore                                             & MLPC                                    & MLPC                                 &  &  \\ \cline{1-3}
zscore                                             & DTW                                     & DTW                                  &  &  \\ \cline{1-3}
mean                                          & L2                                      & Mean                                 &  &  \\ \cline{1-3}
\end{tabular}
\caption{Summary of transformations and distances for our analysis.}
\label{table:distances}
\end{table}

Figure \ref{fig1} shows, for each considered distance, the distribution of the distances for the two groups of series described above. We can observe that, when using the mean, time series from group A are closer to the the reference series than series from group B. That is, mean temperatures in the reference times series are more similar along different scenarios than along geographical points in the same year. But, when considering distances that take into account the shape of the series the situation changes drastically. 
Figure \ref{fig1} shows that Haar level 4 already has a similar distance distribution than Haar level 10, revealing that reducing to level 4 is sufficient for our purposes.

\begin{figure}[H]

\begin{subfigure}{0.5\textwidth}
\includegraphics[width=0.99\linewidth,]{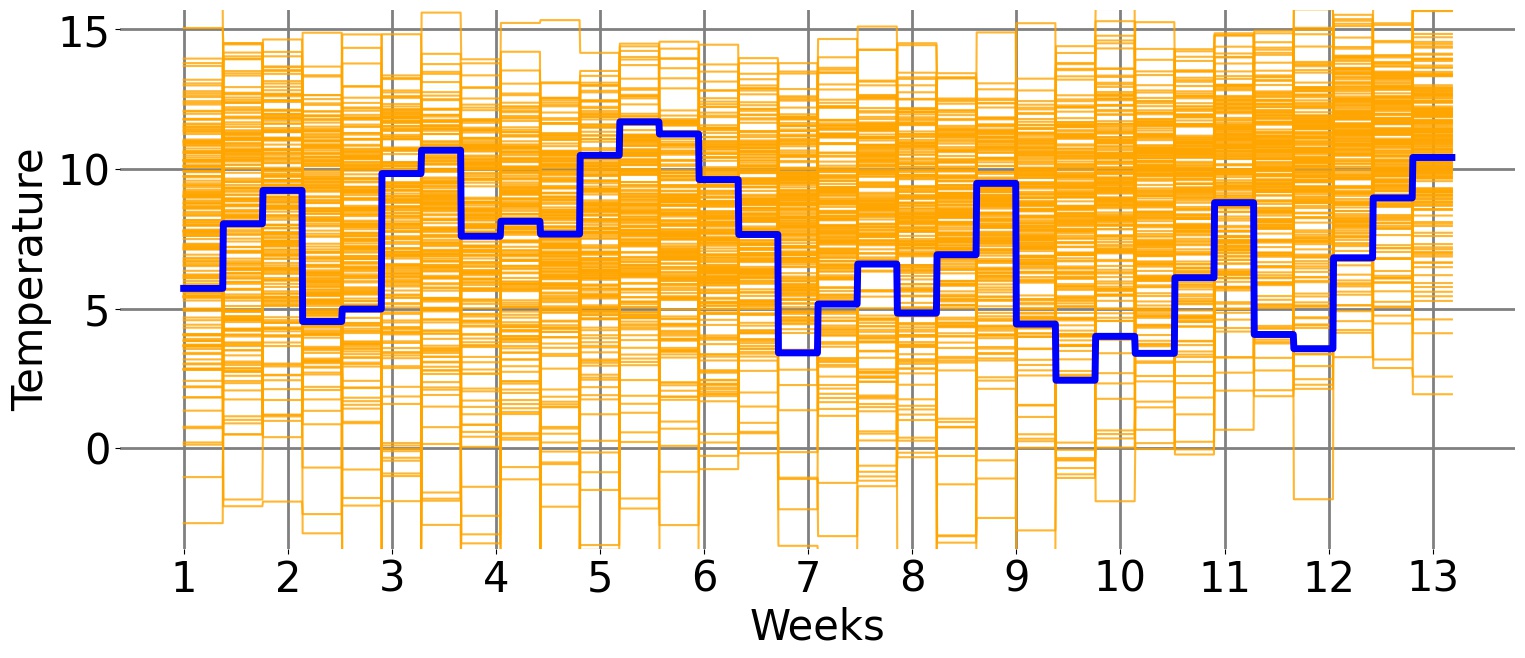} 
\caption{}
\label{fig3a}
\end{subfigure}
\begin{subfigure}{0.5\textwidth}
\includegraphics[width=0.99\linewidth, ]{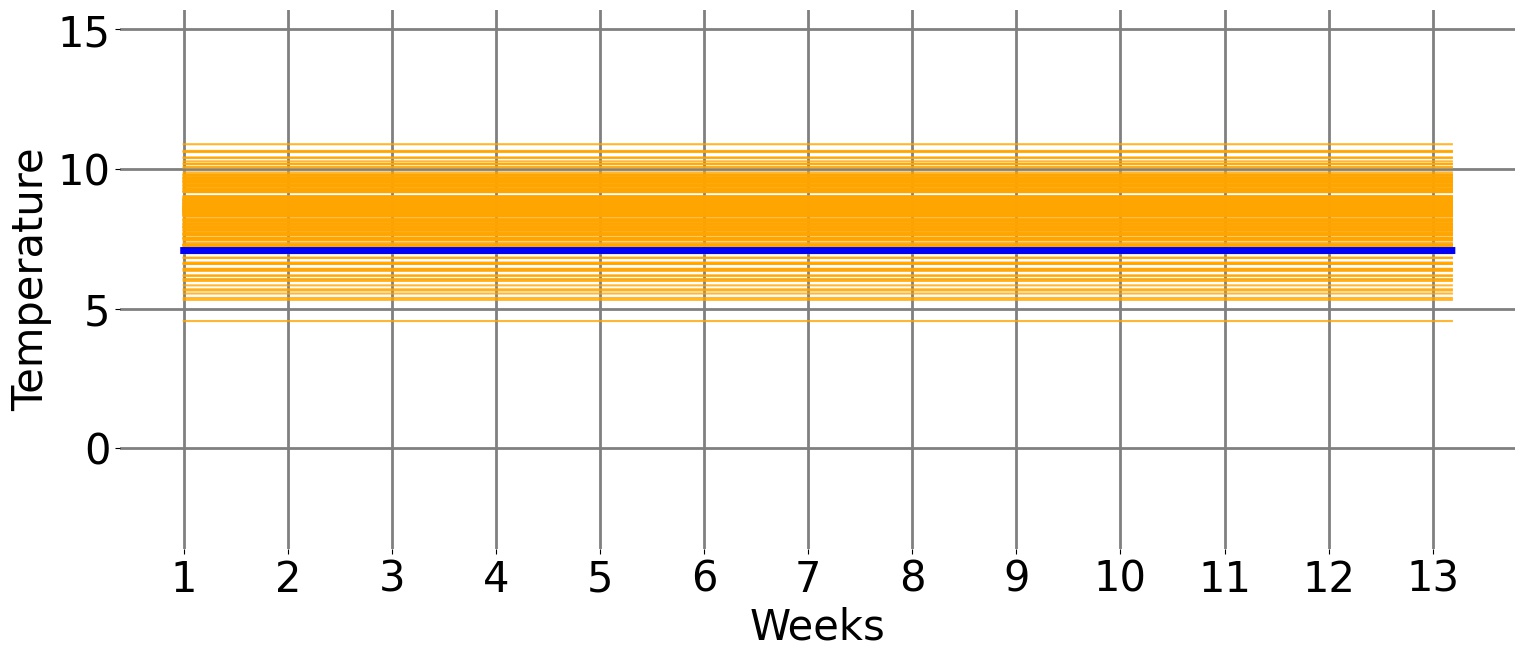}
\caption{}
\label{fig3b}
\end{subfigure}

\begin{subfigure}{0.5\textwidth}
\includegraphics[width=0.99\linewidth,]{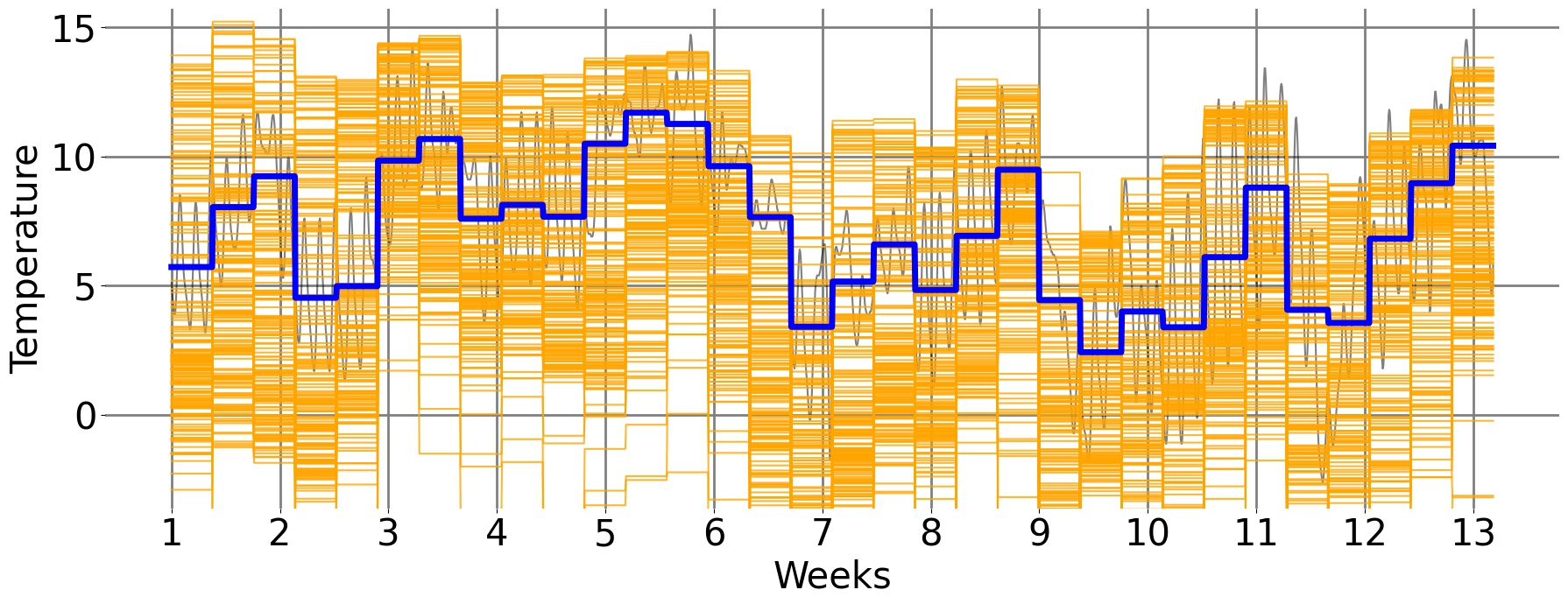} 
\caption{}
\label{fig3c}
\end{subfigure}
\begin{subfigure}{0.5\textwidth}
\includegraphics[width=0.99\linewidth, ]{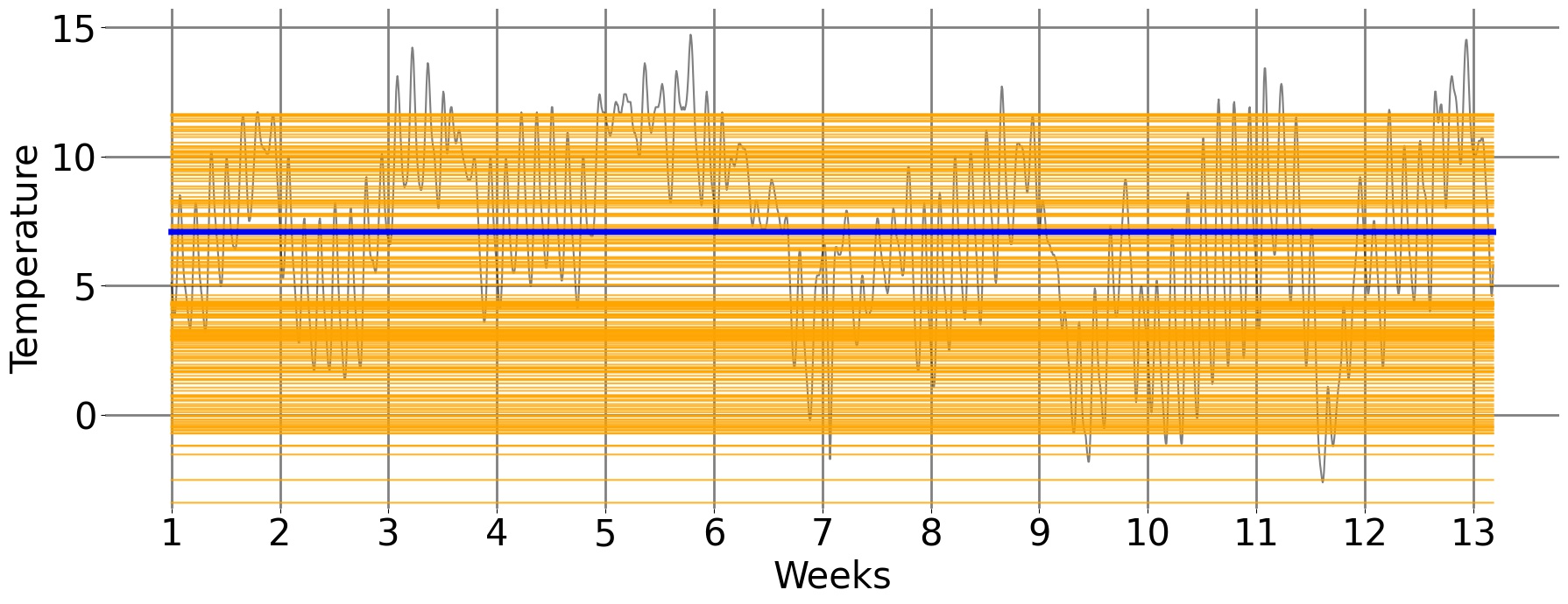}
\caption{}
\label{fig3d}
\end{subfigure}

\caption{Comparison between times series in group A and B by its Haar transformation level 5 (Figures (a) and (c)) and mean temperature (Figures (b) and (d)). Reference time series (without any transformation) in gray, reference time series transformed by Haar level 5 in blue  in (a) and (c).}
\label{fig3}
\end{figure}

The blue distributions progressively shift to the right of the orange distributions, which means that distance of the series from group B to the reference one become progressively smaller than the ones from group A. In order to understand this behaviour, Figure \ref{fig3a}  (upper left) displays the reference time series in gray while Figure \ref{fig3b} displays in orange the Haar level 5 representation of the times series corresponding to group A (same geographical point, different scenarios).

In \ref{fig3b}, the mean of these time series is plotted in orange. Notice that although the shape of the representations are seemingly quite different, the mean temperatures are close to each other.

In \ref{fig3c}, the same reference time series is displayed in gray, while the transformed Haar level 5 times series corresponding to group B (same scenario, different geographical points) are displayed in orange. In \ref{fig3d}, we show their corresponding means. In this case, we observed that the means are further away from the mean of the reference series but at the same time, curves from \ref{fig3} seem to have similar shapes.

We have hence underlined that
the different distances considered lead to take into account different aspects of the temperature times series. Moreover, from this analysis, we evidenced that the climate scenarios are  diverse in terms of shape (i.e., different scenarios display different shape for the same geographical point) while preserving the mean temperature (the same geographical point displays similar mean temperature along many scenarios). 

In order to be able to understand the transition observed in figure \ref{fig1}, it is crucial to understand the characteristics of each of the considered distances. 

\begin{figure}[H]
    \centering
    \includegraphics[scale=0.35]{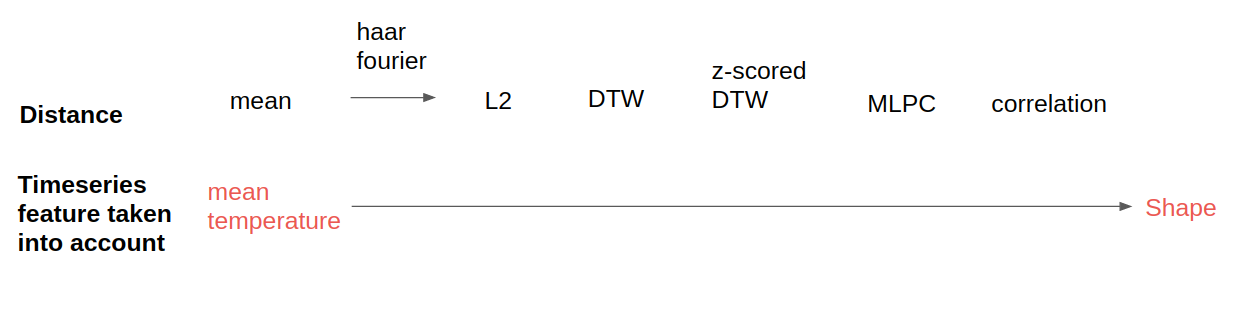}
    \caption{ Distances ordered taking into account how much the shape of the time series is taken into account. The order is schematic and takes into account similarities between the different metrics.}
    \label{fig4}
\end{figure}

In Figure \ref{fig4}, the distances are sorted according to which feature they highlight. On the left side, we consider the mean temperature, and on the right, the correlation.
Keep in mind that Fourier and Haar representation, when having very high level of thresholding, agree with the mean, while
they boil down to L2 in absence of thresholding. 

%% file: secs/experiments.tex
\subsection{Clustering time series for a fixed geographical location}
\label{sec:experiments}

\begin{figure}[H]
 \centering
 \includegraphics[width=0.9\textwidth]{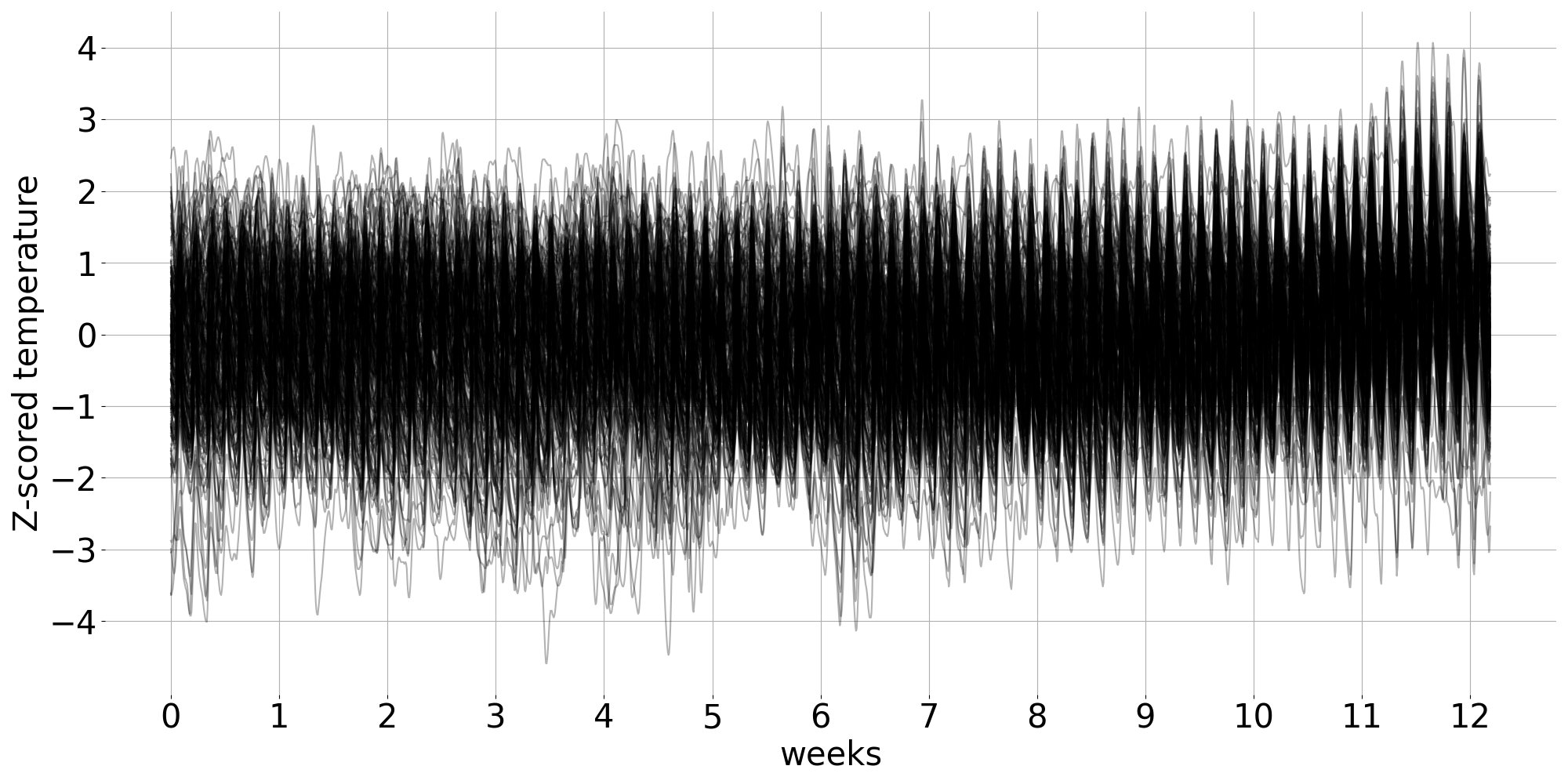}
 \caption{Z-normalized temperature times series corresponding to all the 200 climate scenarios.}
 \label{fig:all_series}
\end{figure}

We now discuss the practical choice of a distance and the resulting clustering
using the previously explained methodology.
In the sequel, we fix one geographical point and consider all the temperatures times series corresponding to the different scenarios for that point. In figure \ref{fig:all_series}, all the  z-normalized temperatures series corresponding to the chosen geographical point are displayed and it can be seen that there is no clear structure, except for the day-night variability. 

Table \ref{tab:models} shows the different models we tried. Each model consists of a representation of the data, a distance and a clustering algorithm. 

\begin{table}[H]
\begin{tabular}{|l|l|l|l|}
\hline
\multicolumn{1}{|c|}{\textbf{Representation of data}} & \multicolumn{1}{c|}{\textbf{Distance}} & \multicolumn{1}{c|}{\textbf{Clustering method}} & \multicolumn{1}{c|}{\textbf{Name}} \\ \hline
mean                                                  & L2                                     & k-medoids                                       & mean                               \\ \hline
plain time series                                                     & L2                                     & k-medoids                                       & L2                                 \\ \hline
Fourier                                               & L2                                     & k-medoids                                       & Fourier                            \\ \hline
Haar                                                  & L2                                     & k-medoids                                       & Haar                               \\ \hline
PCA                                                   & L2                                     & k-medoids                                       & PCA                                \\ \hline
zscore                                                     & max lagged pearson correlation         & k-medoids                                       & MLPC                                \\ \hline
zscore                                                & DTW                                    & k-medoids                                       & DTW                                \\ \hline
\end{tabular}
\caption{ Models.}
\label{tab:models}
\end{table}

Each representation and distance is able, in principle, to capture different features of the data. 
Representing a scenario with its annual mean is an oversimplification but is useful as a baseline for comparisons. PCA95, Fourier95 and Haar95 rely on a dimensional reduction, PCA based on the data itself and Fourier and Haar in the respective bases. On the other hand, MLCC and DTW are able to capture similarities regardless of shift and scale translations.  
We kept the clustering method as simple as possible and chose k-medoids for all cases.\\

We first give evidence that consensus and
within index are a poor overall performance choice in this context. Indeed, it gives a great advantage to drastic dimensional reduction, loosing the patterns that one is looking for a in the data.

\begin{figure}[H]
\centering
\includegraphics[scale=0.37]{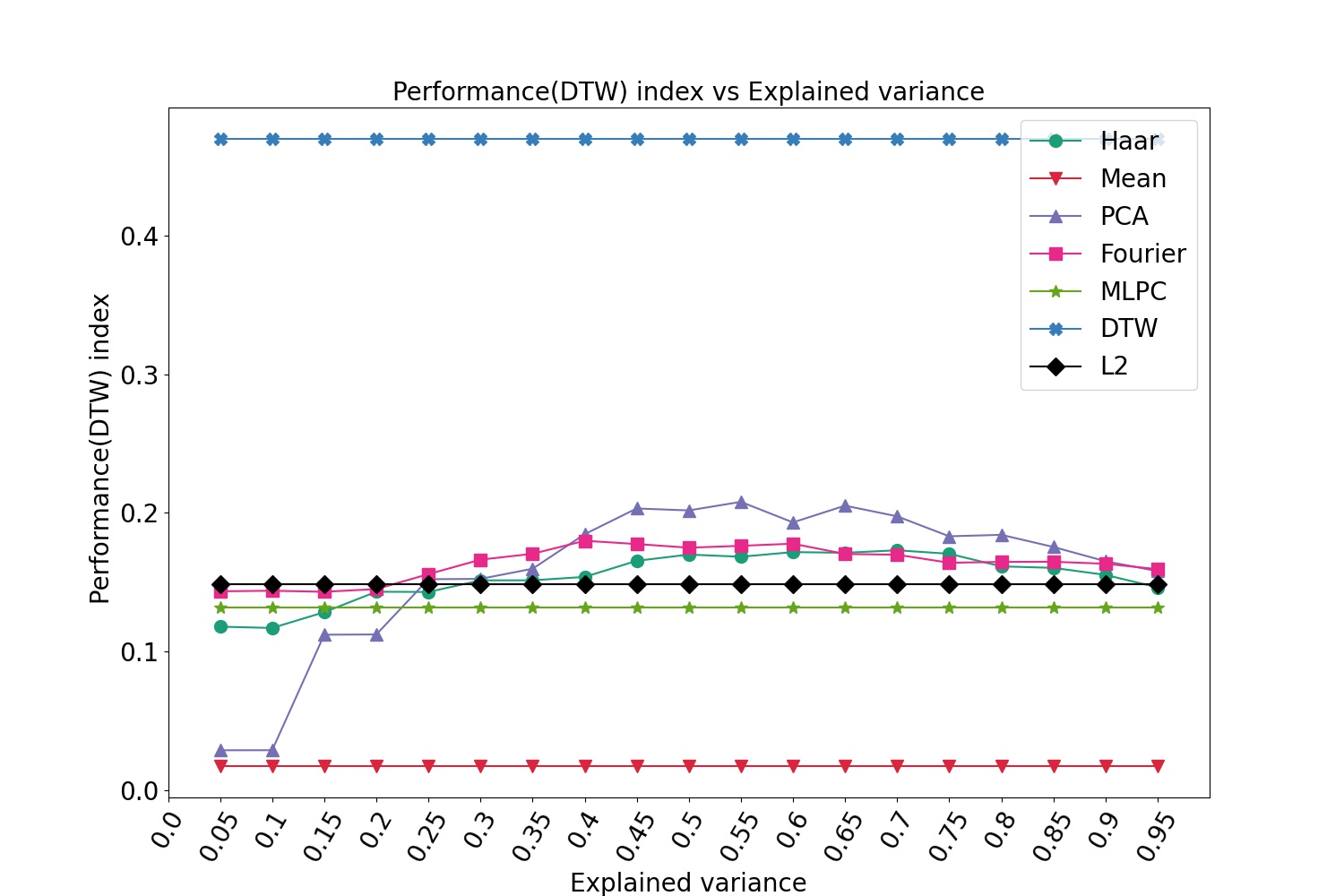}
\caption{Performance index when the reference metric is DTW.}
\label{fig:MLCC}
\end{figure}

\begin{figure}[H]
\centering
\includegraphics[scale=0.37]{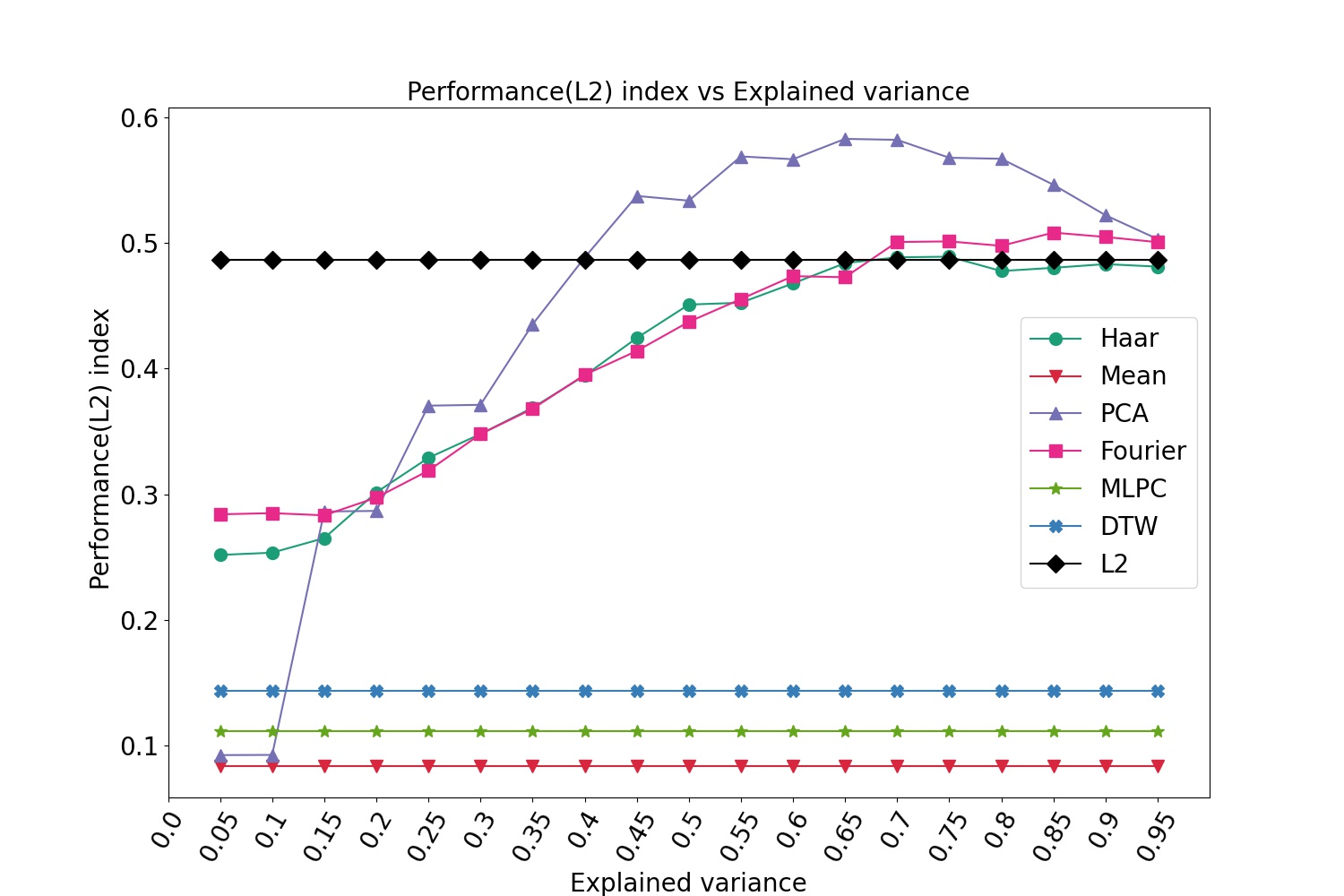} 
\caption{Performance index when the reference metric is L2.}
\label{fig:l2}
\end{figure}

\begin{figure}[H]
\centering
\includegraphics[scale=0.37]{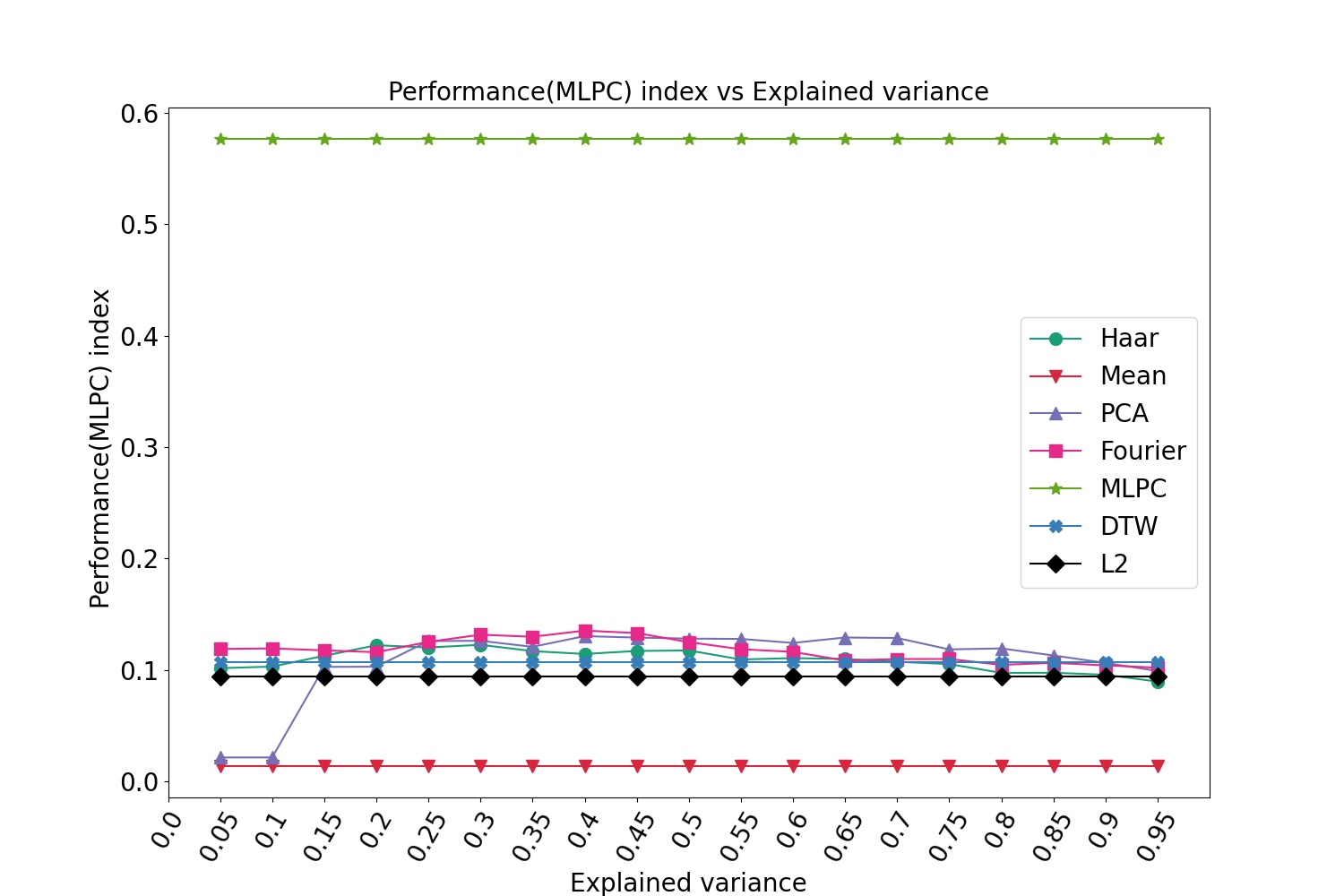} 
\caption{Performance index when the reference metric is MLPC.}
\label{fig:dtw}
\end{figure}

Focusing on our new definition of performance index, we consider the following experiments.
Given that k-medoids uses random points for the initial values of the centers,
we perform several clustering runs (that is, run 5 times the corresponding clustering algorithm) and compute the mean index between the labeled outputs.
Figure \ref{fig:MLCC}, \ref{fig:l2}, \ref{fig:dtw} show the index when using different metrics as references.

In this work, we did not focus on the important problem of selecting the number of clusters and simply set the number of clusters to 15, based on domain experts advice.
Note however, that we performed the same experiments for different numbers and concluded our analysis and results stay valid for a range of cluster numbers close to 15.

While the consensus index does not give enough evidence to select any model,
(and would lead eventually to choose the representation keeping only the means of the time series), our index allows to order conveniently the dimension reductions techniques and to naturally define a best one.

It is remarkable to note that in case of the  underlying distance being max-lagged correlation or DTW, then
this distance (without any dimension reduction on the signals) is clearly 
the best choice. In future work, it would be interesting
to define new versions of both distances
on modified versions of the time series incorporating a certain amount of dimension reduction.
In the  case of L2 however, reducing the dimension clearly pays off and there is a non trivial optimum.

\begin{figure}[H]
\centering
\includegraphics[scale=0.37]{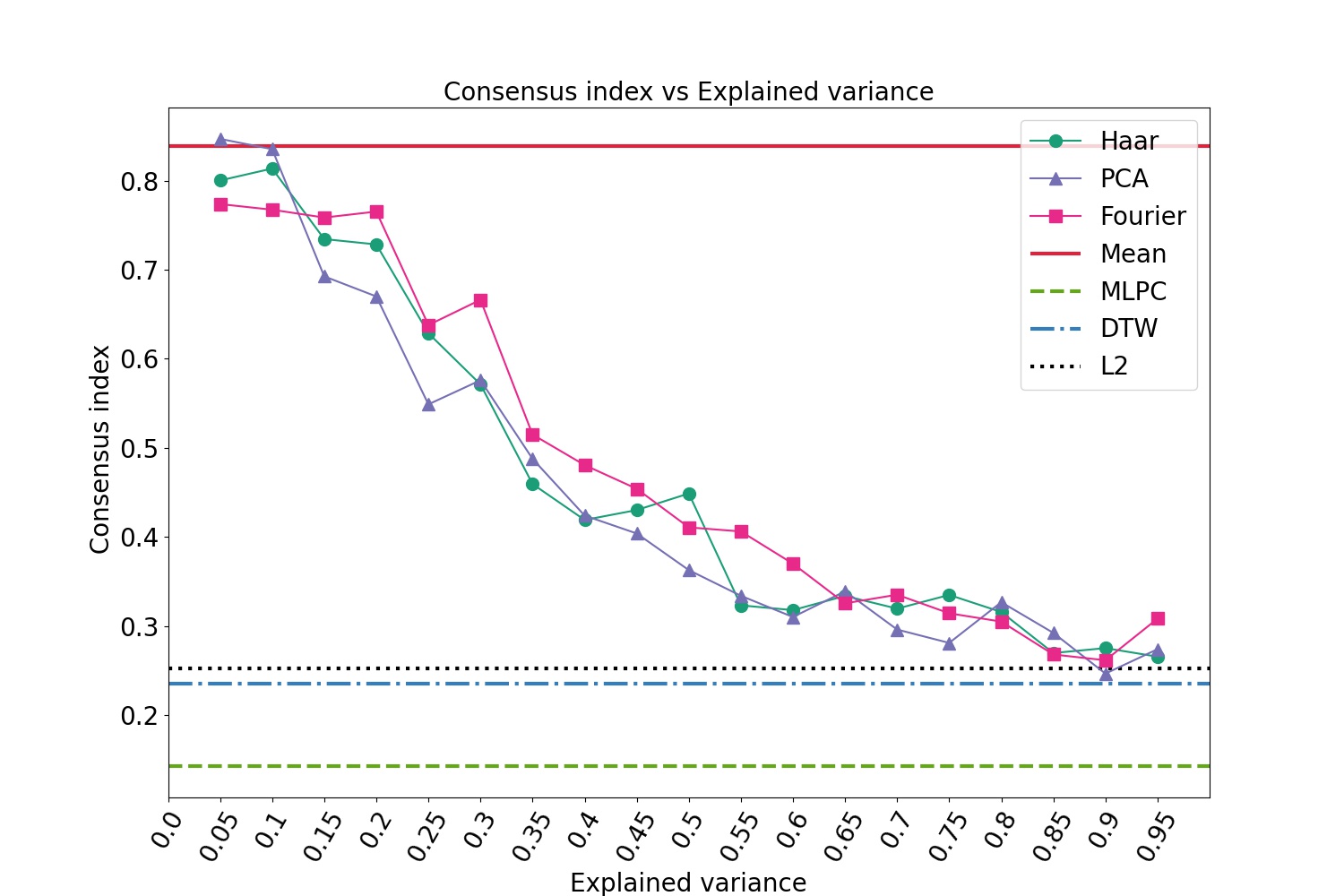}
\caption{Consensus indexes for the different metrics considered.}
\label{fig:consensus}
\end{figure}

\begin{figure}[H]
\centering
\includegraphics[scale=0.37]{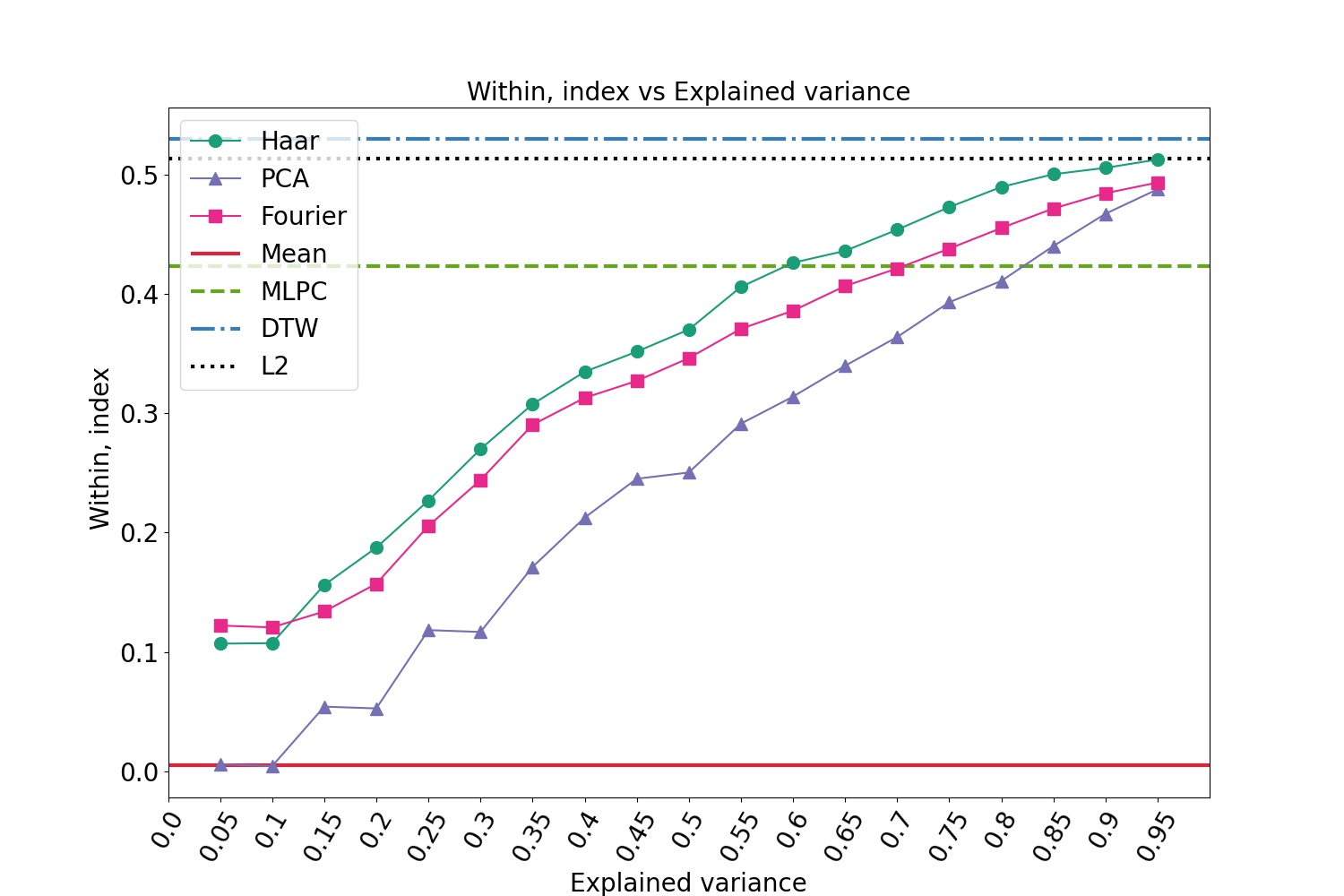} 
\caption{Within index for the different metrics considered.}
\label{fig:within}
\end{figure}

\subsubsection{Description of the clusters}

\begin{figure}[H]
   \centering
  \includegraphics[width=.85\textwidth]{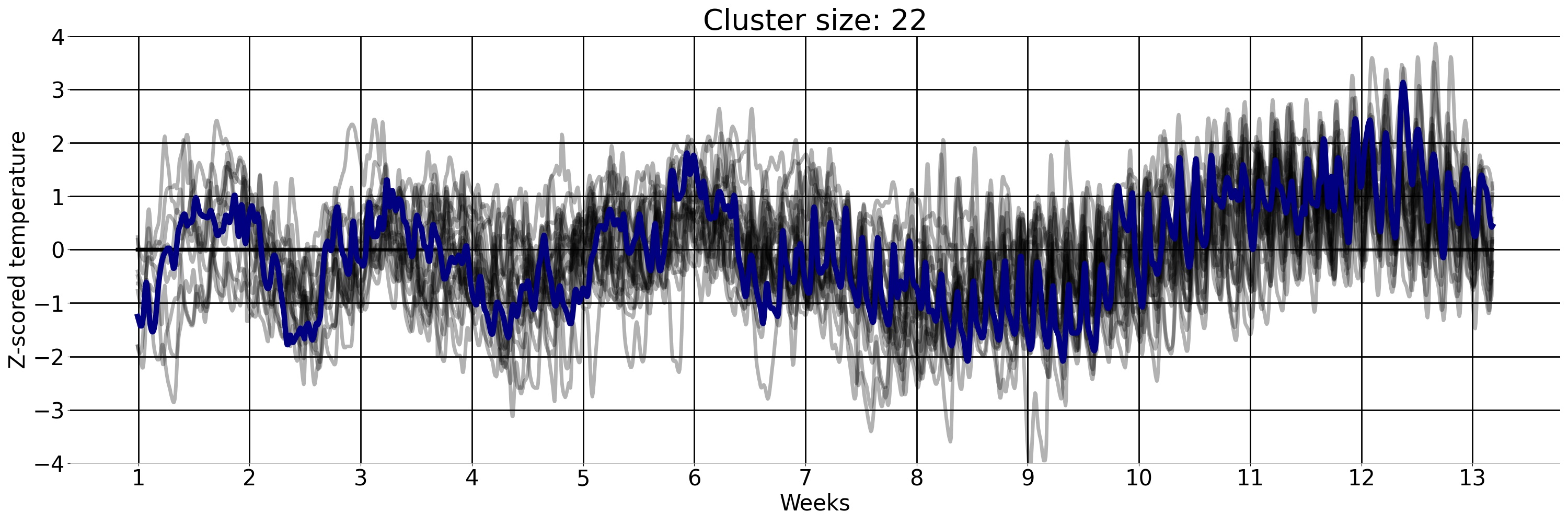}
   \includegraphics[width=.85\textwidth]{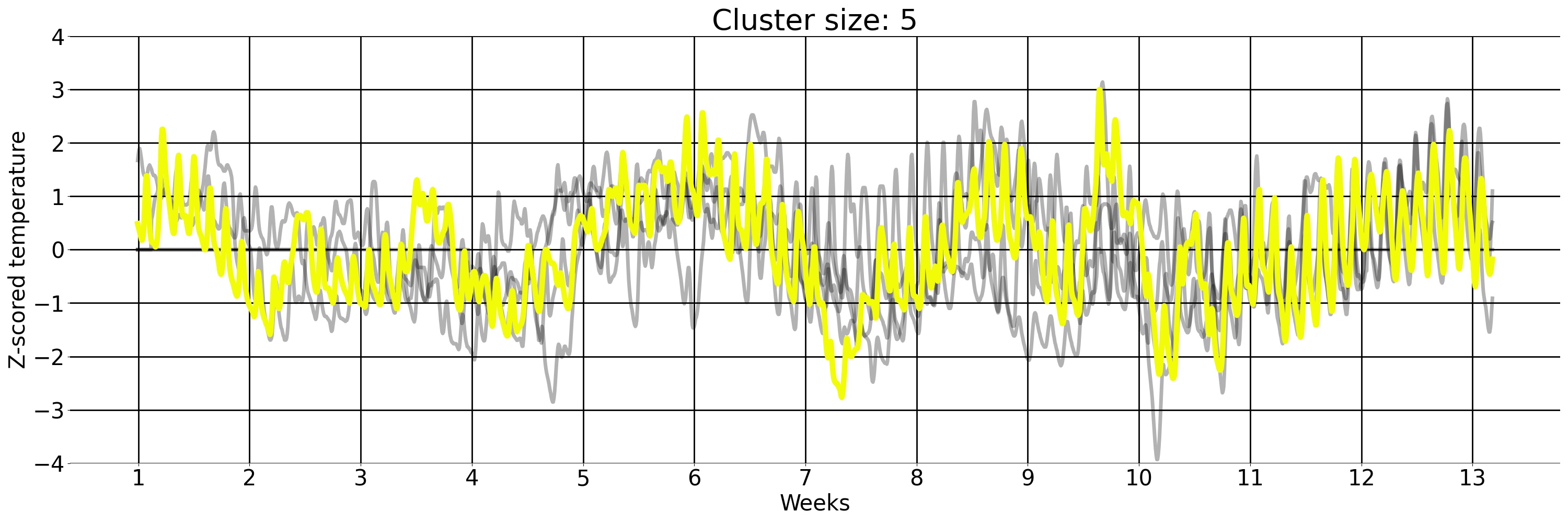}
   \caption{ Series in the cluster in gray and the cluster representative in color. MLPC metric.}
   \label{fig:MLPC-clusters-1}
\end{figure}

\begin{figure}[H]
   \centering
   \includegraphics[width=.85\textwidth]{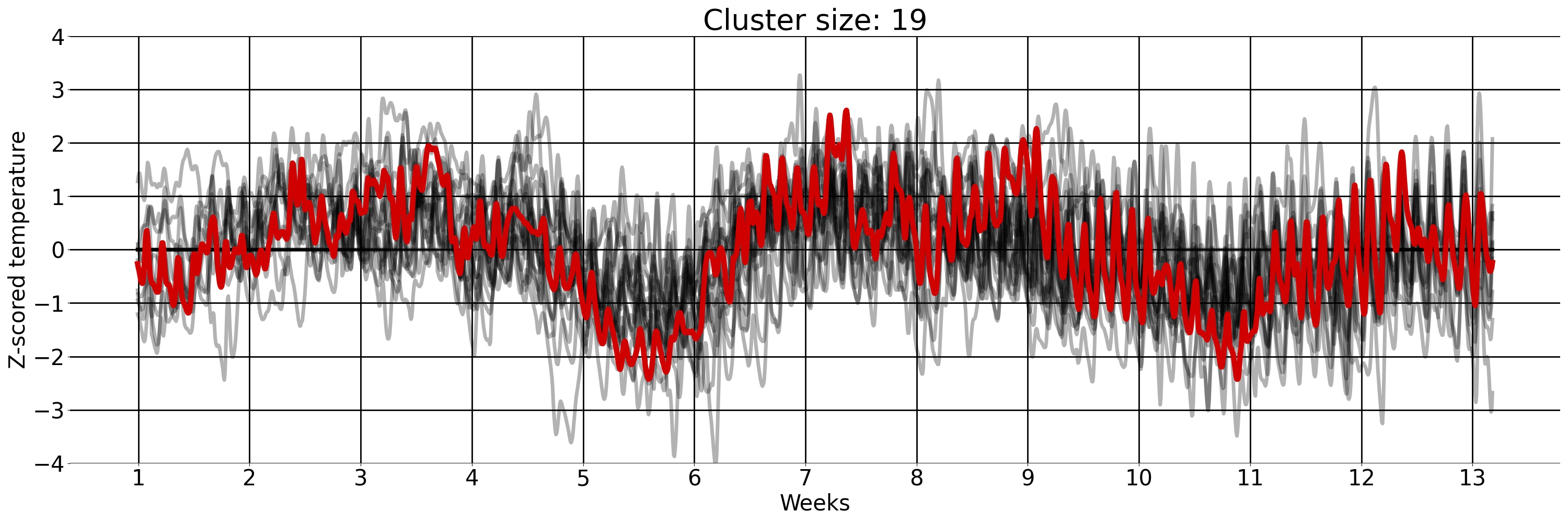}
   \includegraphics[width=.85\textwidth]{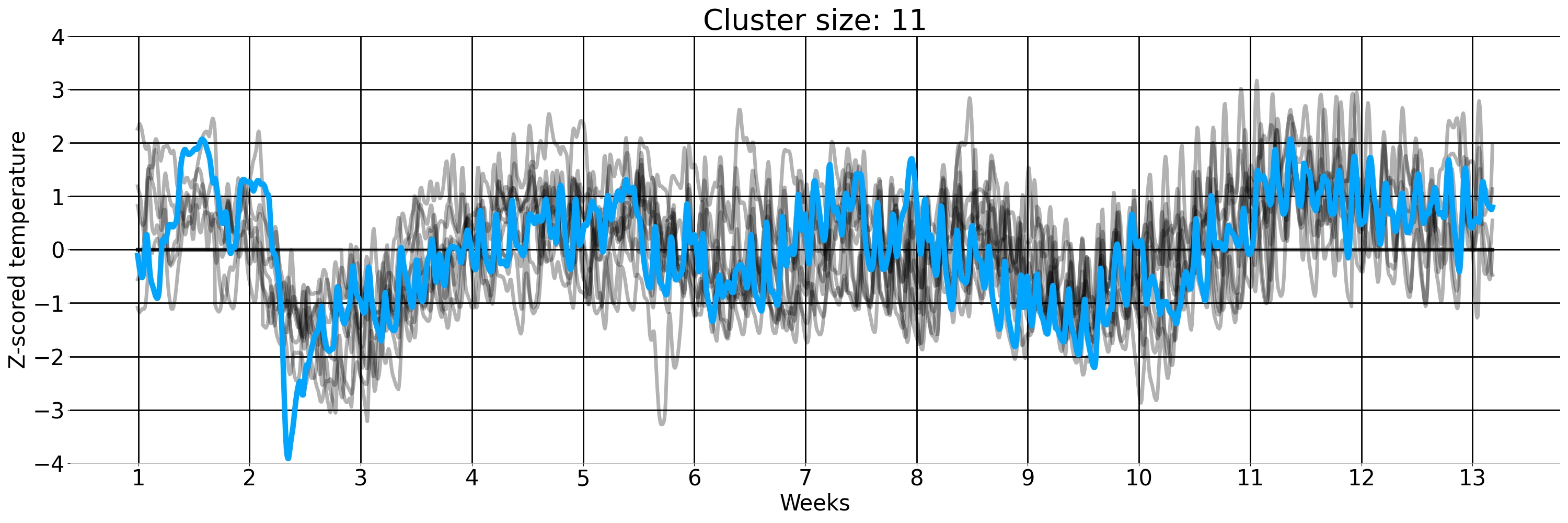}
   \caption{ Series in the cluster in gray and the cluster representative in color. MLPC metric.}
   \label{fig:MLPC-clusters-2}
\end{figure}

\begin{figure}[H]
   \centering
  \includegraphics[width=.85\textwidth]{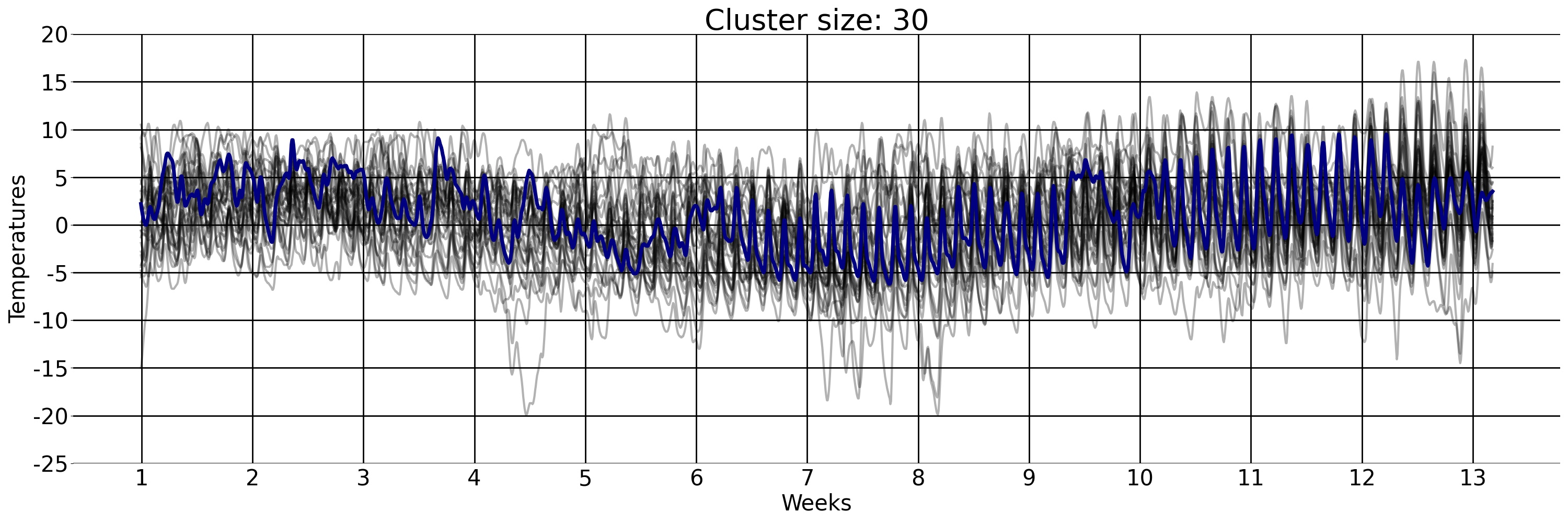}
   \includegraphics[width=.85\textwidth]{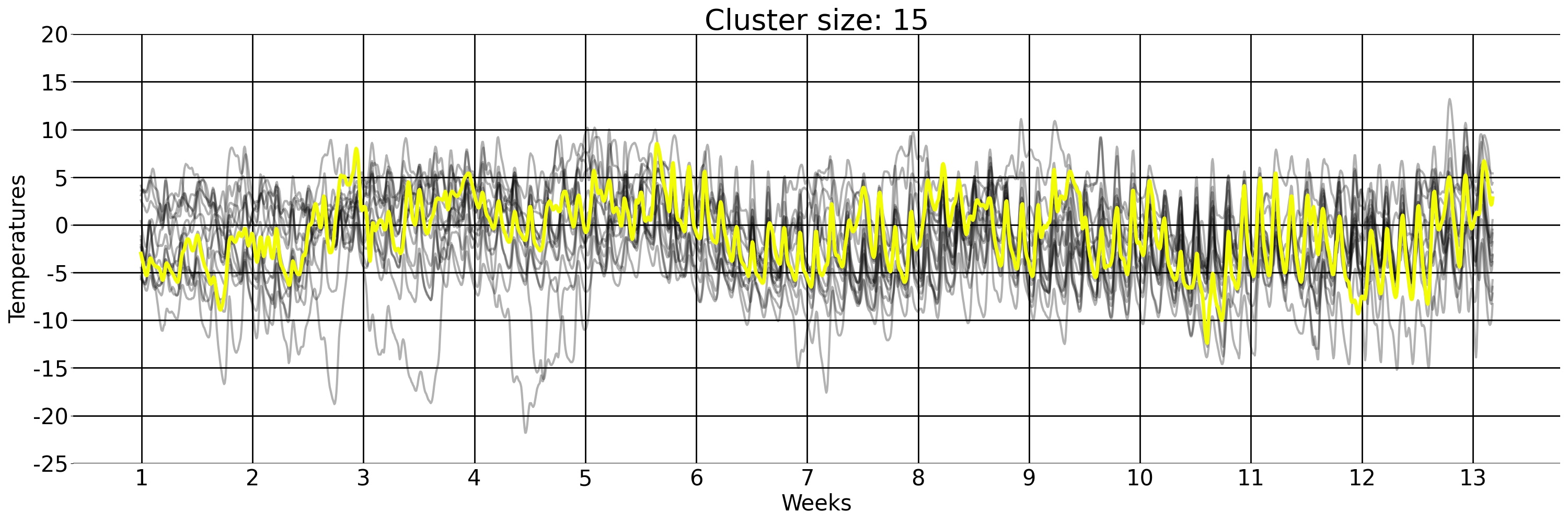}
   \caption{ Series in the cluster in gray and the cluster representative in color. PCA95 metric. }
   \label{fig:PCA-clusters-1}
\end{figure}

\begin{figure}[H]
   \centering
   \includegraphics[width=.85\textwidth]{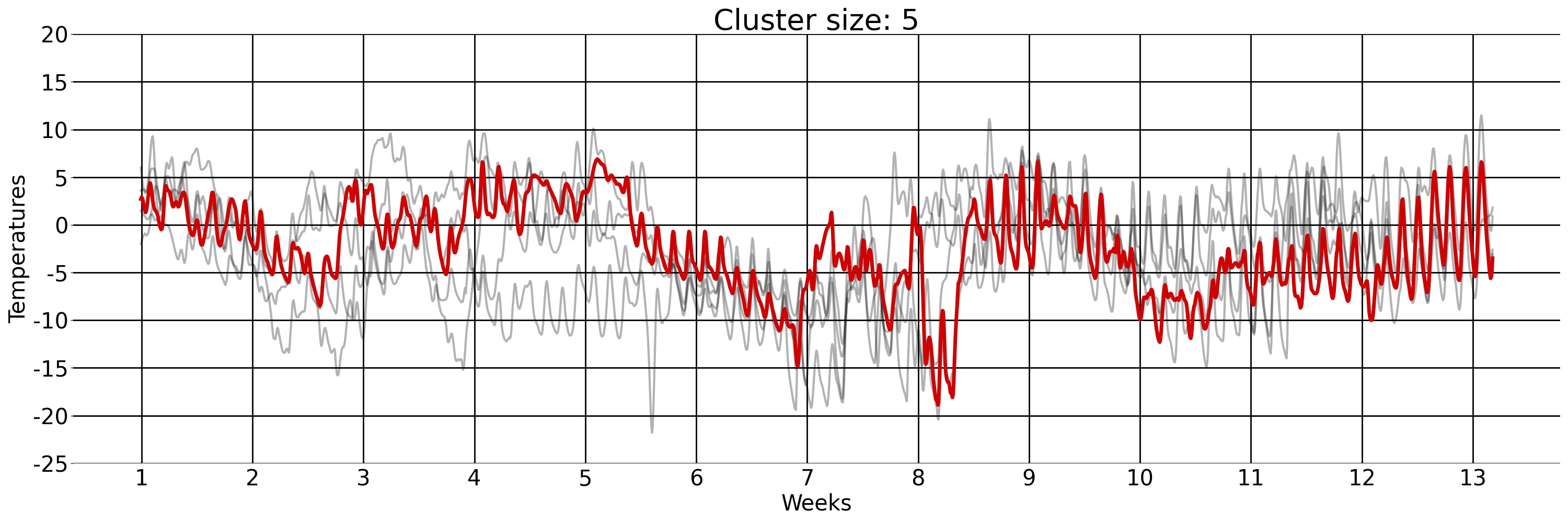}
   \includegraphics[width=.85\textwidth]{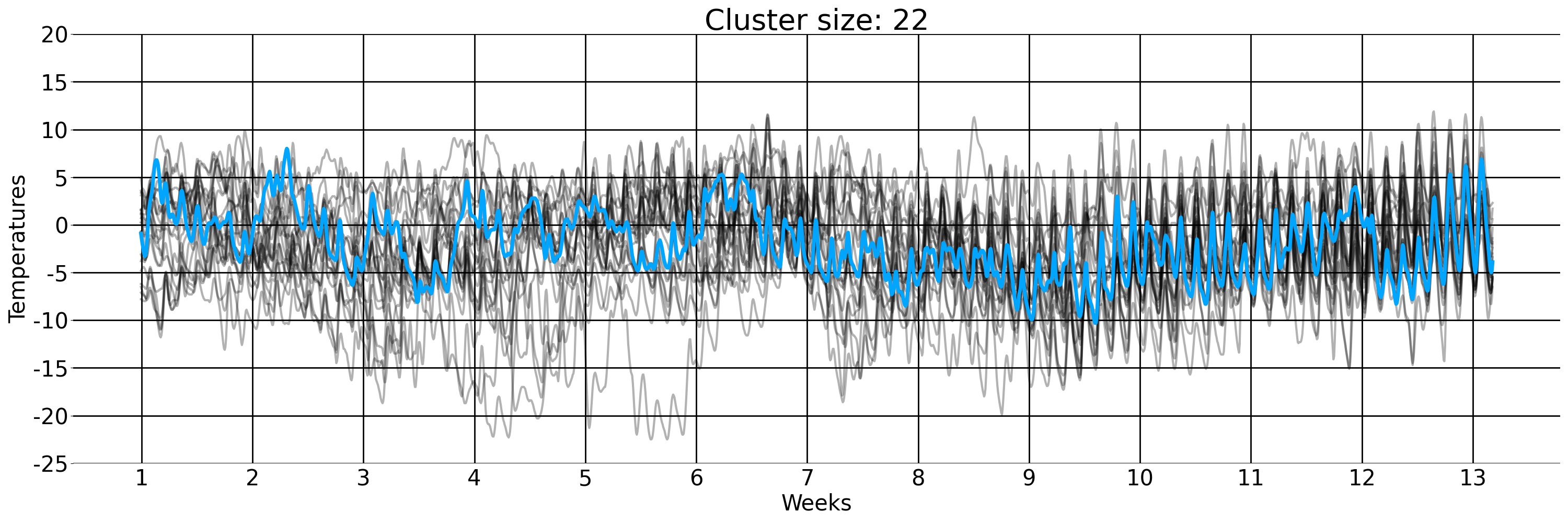}
   \caption{ Series in the cluster in gray and the cluster representative in color.PCA 95 metric.}
   \label{fig:PCA-clusters-2}
\end{figure}

Figure \ref{fig:MLPC-clusters-1} and \ref{fig:MLPC-clusters-2} shows four resulting clusters for the MLPC model out of 15 clusters. Figure \ref{fig:PCA-clusters-1} and \ref{fig:PCA-clusters-2} shows four resulting clusters out of 15 for the PCA95 model out of 15 clusters. 

In each cluster, the cluster representative is in color (or thick black) and  the other series that belong to that cluster are in gray. The representative is the 
barycenter of the cluster series. All the series in the cluster are lagged to maximize the correlation with the cluster barycenter. 
It can be clearly seen that there is common pattern among the series in each cluster and that the cluster representative in indeed reflecting the overall behavior of the cluster.

%% file: secs/conclusion.tex
We explored the clustering of temperature time series and we showed that the distance employed by the clustering is crucial in terms of results interpretation and should be chosen with case in function of the spatio/temporal analysis undertaken.
We then focused on clustering scenarios for a given location and showed that given a distance, well performing clustering
can be chosen through maximizing an index balancing the trade-off between clustering efficiency and dimension reduction.
Building on these conclusions, it would be interesting
to revisit the loss functions proposed for deep clustering where
the dimension reduction and the clustering are considered jointly.